\documentclass[twocolumn,twoside,slac_two]{revtex4}
\usepackage{graphicx}
\usepackage{fancyhdr}
\usepackage{hyperref}
\hypersetup{
    colorlinks=true,
    urlcolor=magenta,
}
\begin{document}

\title{LHC results and hadronic interaction models}

\author{S.\ Ostapchenko}
\affiliation{Frankfurt Institute for Advanced Studies,
 60438 Frankfurt am Main, Germany}
\affiliation{D.V. Skobeltsyn Institute of Nuclear Physics,
Moscow State University, 119992 Moscow, Russia}

\begin{abstract}
The present status of high energy cosmic ray interaction models
is discussed, concentrating on recent model updates inspired
 by the data from Run 1 of the LHC.
A special attention is devoted to the remaining differences in the model
predictions and their relation to the underlying theoretical approaches.
Opportunities for the model discrimination by future LHC and cosmic ray
experimental studies are analyzed.
\end{abstract}

\maketitle

\thispagestyle{fancy}

\section{INTRODUCTION}
\label{sec:intro}
Modeling of high energy hadronic interactions is of considerable importance
for experimental studies of very high energy  cosmic rays (CRs), especially,
for an analysis of the primary CR composition. 
Applying the  traditional extensive air shower (EAS) techniques, i.e.\ 
inferring the properties of the  primary CR particles  from measured 
characteristics of nuclear-electro-magnetic cascades induced by their
interactions in the atmosphere, the different  primaries are discriminated
based on the respective differences between some basic air shower observables,
which in turn depend on the way those particles interact with air nuclei. 

There are two main experimental procedures \cite{nagano,ung12}.
When studying  the longitudinal EAS development by measuring fluorescence
light produced by excited air molecules,
  the  primary  particle type may be inferred from the measured
position of the  shower maximum $X_{\max}$ -- the depth in the
atmosphere (in g/cm$^2$) where the number of ionizing particles reaches its
maximal value. Remarkably,  $X_{\max}$ depends strongly on the
properties of the primary particle interaction  with air nuclei, notably, on
the inelastic cross section and the forward spectra of secondary hadrons
produced.  Thus, here one may expect to profit maximally
from experimental studies of proton-proton and proton-nucleus collisions at the
Large Hadron Collider (LHC).
On the other hand, when studying air showers   with scintillation
detectors positioned at ground,  the primary particle type is inferred from
 the relative fraction of muons, compared to all charged particles at ground.
As  the EAS muon content is formed in a multi-step cascade
process, driven mostly by interactions of secondary pions and kaons with air,
the muon density   $\rho_{\mu}$ at ground depends strongly on the properties
 of pion-air collisions over a wide range of energies. In particular,
  any searches
 for new physics signals with ground-based EAS detectors should be very
 challenging since the sensitivity to the
properties of the primary particle interaction is considerably weakened
by the cascade development.

In the following,  we are going to discuss the impact of LHC measurements
on the modeling of high energy cosmic ray interactions, concentrating in
particular on the remaining differences between the predictions of the
popular hadronic interaction models.
We shall compare the results of the most recent   versions of the 
  EPOS \cite{wer06,pie15},  
 QGSJET-II \cite{ost06,ost11}, and SIBYLL  \cite{fle94,rie15} models, which 
 all have  been updated using   experimental data from Run 1 of the LHC.
  Additionally,
 we shall use the QGSJET model  \cite{qgs93,qgs97} which,
  though being already outdated physics-wise, demonstrated a generally 
good agreement with the LHC data  \cite{den11}. Hence, it will be used here
 to study the range of potential variations of model predictions for $X_{\max}$,
 in view of current LHC data.
 Our primary goals are to analyze  the differences between the model results, 
 to trace their origin to the underlying approaches for the treatment of 
 hadronic collisions, implemented in those models, and to propose potential 
 measurements by LHC and cosmic ray experiments, which could be able 
 to discriminate between the alternative approaches.

\section{IMPACT OF LHC DATA AND REMAINING   DIFFERENCES BETWEEN THE MODEL
PREDICTIONS}
\label{sec:model-xmax}
For all  contemporary Monte Carlo generators of high energy hadronic
 collisions,
the qualitative picture behind is the one of quantum chromodynamics (QCD):
The interactions are mediated by multiple cascades of partons (quarks
and gluons) developing between the projectile and target hadrons or nuclei.
There comes the predictive power of the models: Ones the
treatment of the interaction mechanism is developed and the respective
parameters  are fixed, based on some set of experimental data, a particular
model is able to predict the interaction properties at a higher energy or
in a different kinematic range. In particular, changing from proton-proton
interaction to the pion-proton case or to proton-nucleus (nucleus-nucleus)
collisions implies a change of the initial conditions for those parton
cascades, without changing the interaction mechanism itself. Yet the 
corresponding treatments are largely based on phenomenological approaches:
While the perturbative QCD allows one to describe the evolution of ``hard''
(high transverse momentum $p_{\rm t}$) partons, it is of little help for many
other important aspects, like the evolution of ``soft'' (small $p_{\rm t}$) 
partons, the multiple scattering mechanism, and the very initial conditions
for the parton cascades. Therefore, new experimental data corresponding to a
different energy or kinematic range are very valuable for tuning the parameters
of such phenomenological models and, more importantly, for discriminating
invalid theoretical solutions.

\begin{figure}[t]
\centering
\includegraphics[height=5.9cm,width=0.45\textwidth,clip]{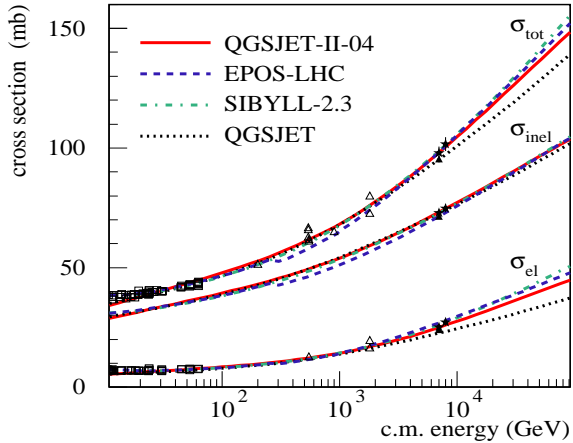}
\caption{$\sqrt{s}$-dependence of the total, inelastic, and elastic $pp$
 cross sections, as calculated using the QGSJET-II-04 \cite{ost11},
  EPOS-LHC \cite{pie15}, SIBYLL-2.3 \cite{rie15},
 and QGSJET  \cite{qgs97} models (solid, dashed, dash-dotted, and dotted 
 lines respectively).
Experimental data are from Refs.\  \cite{totem13,aad14,pdg}.}
\label{fig:sigmas}       
\end{figure}

In what concerns cosmic ray interaction models, the most important results
of Run 1 of the LHC have been  precise measurements of
 the total and elastic proton-proton  cross sections by the TOTEM 
 and ATLAS experiments \cite{totem13,aad14}. Apart from reducing
  drastically the differences between
the respective   model predictions  in the limit of ultra-high energies, 
as illustrated in  Fig.~\ref{fig:sigmas}, %
those experimental results constrained  a number of key parameters
of the models, which impact many other model predictions, e.g.\ for
 secondary particle production.
While measurements of secondary particle production at the central
rapidity region by the ALICE, ATLAS, and CMS  experiments at the LHC
 have not revealed any
serious deficiencies of CR interaction models   \cite{den11}, the corresponding
experimental results contributed to fine-tuning of model parameters.
And the new model versions appeared to be in a reasonably good agreement
with experimental data from LHC Run 2 on soft particle
 production \cite{kha15,aab16,ada16}.

\begin{figure}[htb]
\centering
\includegraphics[height=5.9cm,width=0.45\textwidth,clip]{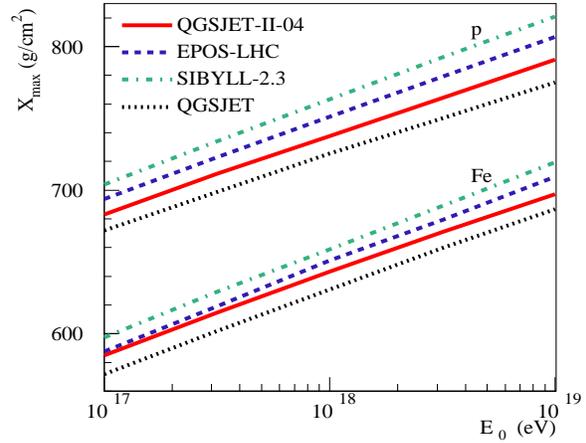}
\caption{Primary energy dependence of the average shower maximum depth 
 for proton- and iron-initiated vertical EAS, as calculated using the 
 QGSJET-II-04, EPOS-LHC, SIBYLL-2.3, and QGSJET models 
 (solid, dashed, dash-dotted, and dotted lines respectively).}
\label{fig:xmax-all}       
\end{figure}

Yet the models diverge considerably in their predictions for EAS properties,
 as illustrated in  Fig.\ \ref{fig:xmax-all} for the particular case of $X_{\max}$. %
 It is thus highly desirable to reveal the reasons for those differences
 and to find ways to further constrain model predictions or, even better,
 to refute some model approaches. In particular, one may hope to gain
insight into the problem, based on measurements of forward hadron
 spectra by the TOTEM and LHCf experiments at the LHC, since the
 corresponding results proved to be a challenge for 
 most of the present Monte Carlo generators
 \cite{cms-tot,ant15,adr11}.

\section{IMPACT OF CONSTITUENT PARTON FOCK STATES}
\label{sec:fock}
\begin{figure}[htb]
\centering
\includegraphics[height=4cm,width=0.23\textwidth,clip]{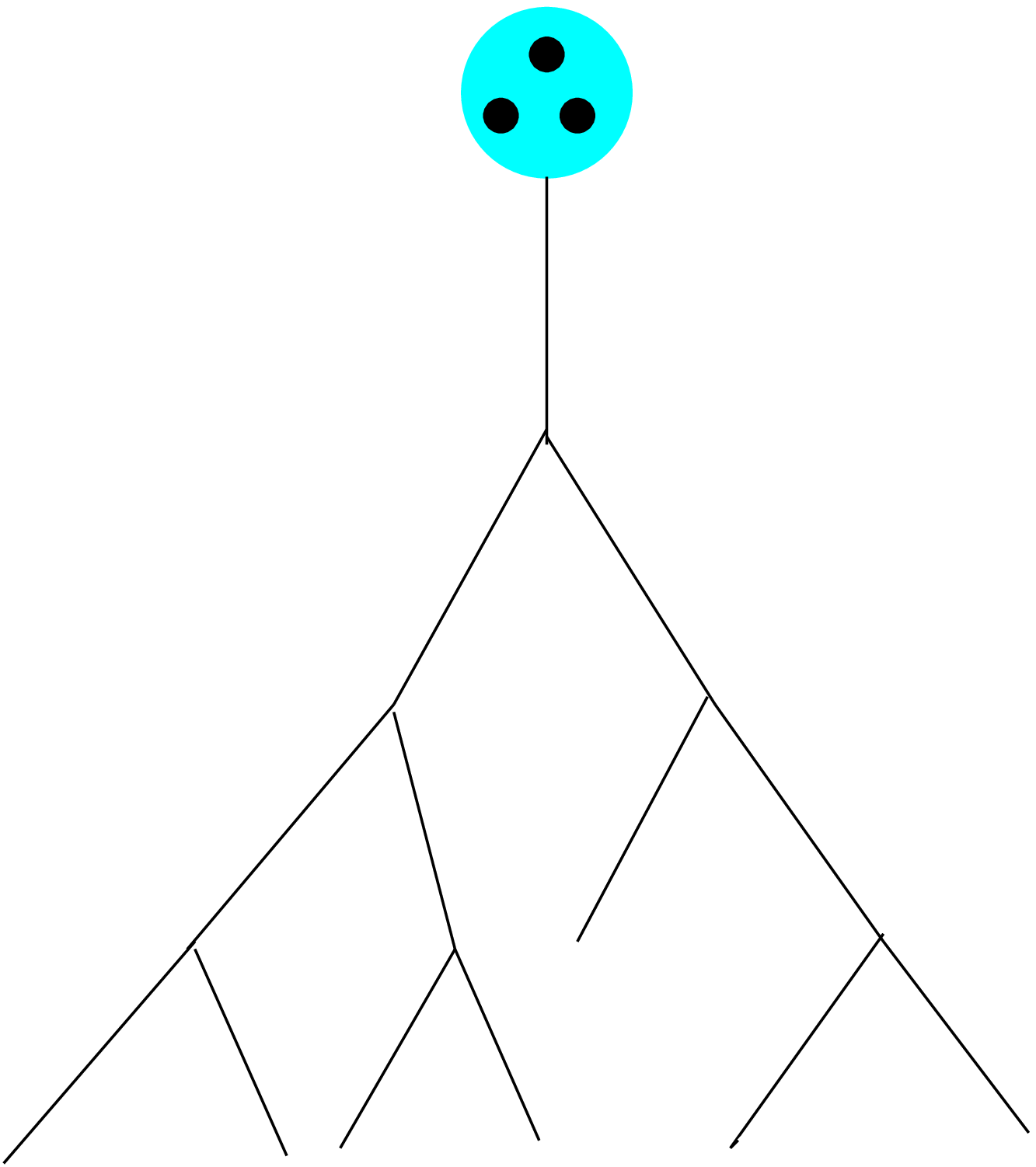}\hfill
\includegraphics[height=4cm,width=0.23\textwidth,clip]{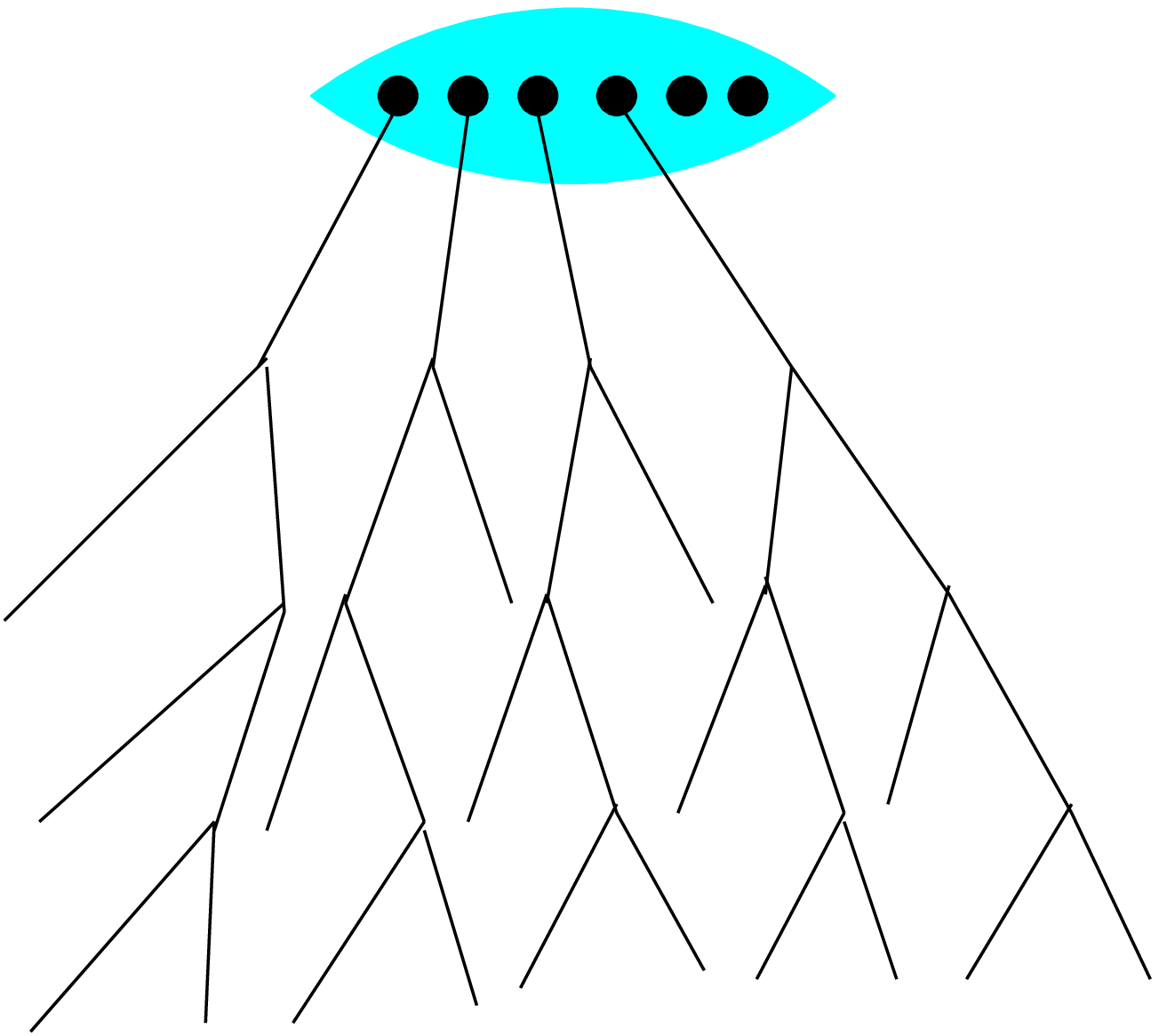}
\caption{Schematic view of the initial part of the parton cascade in the proton.
Left: the cascade starts from the same universal parton Fock state;
new partons participating in multiple scattering processes emerge from the
cascade development, being characterized by $\propto 1/x$  distributions
for the   momentum fraction. Right: the proton is represented by
a superposition of Fock states consisting of different numbers of large $x$
constituent partons; the more abundant multiple scattering the larger
Fock states  involved in the process.}
\label{fig:fock}       
\end{figure}

Let us start with  SIBYLL-2.3  which predicts the largest values
for  $X_{\max}$ and for the shower elongation rate between all the considered
models, as one can see in  Fig.\ \ref{fig:xmax-all}. This appears to be related
 to the very basic model assumptions concerning the structure
of constituent parton Fock states in hadrons, i.e.\ for the above-mentioned
initial conditions for parton cascades, as discussed in more detail
in Ref.\ \cite{ost16}. Like   most of the hadronic  event
 generators used in the collider field,  the SIBYLL model is based
 on the ``minijet'' approach which corresponds 
 implicitly to the picture 
  shown schematically on the left-hand side (lhs) of Fig.\ \ref{fig:fock}.
 At large Feynman $x$, one   starts from the same universal  
parton Fock state.  Additional partons  (sea quarks or gluons)
 giving rise to new branches of the parton cascade, which take
part in the multiple scattering processes, result from the evolution of the
parton density corresponding to this initial 
state and their momentum fractions are  distributed as
$\propto 1/x$ in the very high energy limit.
Such a picture reflects itself in the hadron production pattern predicted
by the model: Multiple scattering affects mostly central particle production,
while having a weak influence of forward hadron spectra. Indeed, the latter
are formed by the hadronization of partons emerging from the
initial part of the underlying parton cascade, which starts from the same
initial conditions and covers a short rapidity interval, 
being thus weakly dependent on the further development of the cascade.

\begin{figure}[b]
\centering
\includegraphics[height=5.9cm,width=0.45\textwidth,clip]{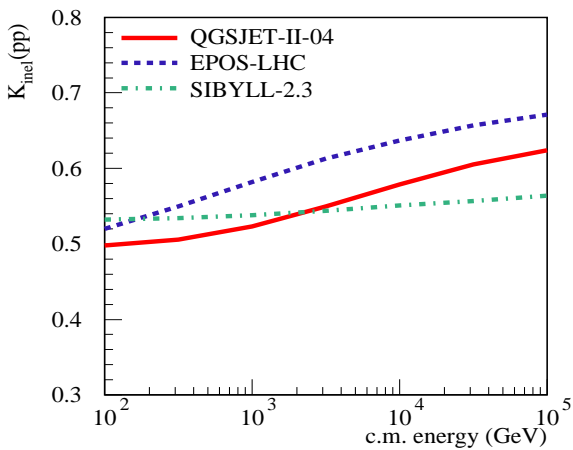}
\caption{Energy dependence of the inelasticity of leading nucleons
 in $pp$ collisions,  as calculated using the 
 QGSJET-II-04, EPOS-LHC, and SIBYLL-2.3 models
 (solid, dashed, and dash-dotted lines respectively).}
\label{fig:inel}       
\end{figure}

A direct consequence of the above-discussed approach is  
a weak energy dependence of  the inelasticity $K^{\rm inel}$, i.e.\
the relative energy loss of leading nucleons, in proton-proton and
proton-nucleus collisions. With increasing energy, one obtains a
significant enhancement of secondary particle production in the central 
rapidity region only, which has a weak impact on the energy loss of 
leading nucleons. As one can see in Fig.~\ref{fig:inel}, %
 the energy dependence of  $K^{\rm inel}_{pp}$ is indeed almost flat 
for SIBYLL-2.3. In turn, a slower energy-rise of the inelasticity implies
 a  larger EAS elongation rate and a larger  $X_{\max}$ at
sufficiently high energies (see, e.g.\  Ref.\ \cite{ulr11}),
as we observed indeed  in  Fig.\ \ref{fig:xmax-all}. 
 
In the alternative approach, implemented in the EPOS and QGSJET(-II) models,
 a proton is represented by a superposition
of a number of Fock states containing different numbers of large $x$ 
constituent partons, as shown schematically on the right-hand side (rhs) of
 Fig.\ \ref{fig:fock}.
 Further cascading of these partons ``dresses''  them with  low $x$ parton
clouds. As the overall parton multiplicity in the central rapidity region
is roughly proportional to the
number of initial constituent partons, stronger multiple scattering is
 typically associated with larger Fock states. Thus, there is a strong
long-range correlation between central and forward particle production; higher
 multiplicity in the central  region reflects stronger multiple 
 scattering. In turn, this implies that bigger numbers of large $x$ 
 constituent partons are   involved  in the process,
  which has a strong impact on forward hadron spectra.

This naturally leads to a substantial energy-rise of the
inelasticity, which is clearly seen   in Fig.~\ref{fig:inel} for 
  QGSJET-II-04 and   EPOS-LHC.
 The reason for this rise is twofold. First, for any given Fock state,
 increasing multiple scattering implies that bigger numbers of large $x$
 constituent partons are involved in the interaction, thus leaving
smaller fractions of the initial proton momentum  for spectator partons
which finally form the leading nucleons. 
Additionally, Fock states with bigger and bigger numbers of large $x$
constituent partons come into play. Momentum sharing between these partons
results in  a smaller fraction of the initial proton momentum, possessed by
each parton, which thus enhances the energy loss of the leading nucleons.

\begin{figure}[t]
\centering
\includegraphics[height=5.9cm,width=0.45\textwidth,clip]{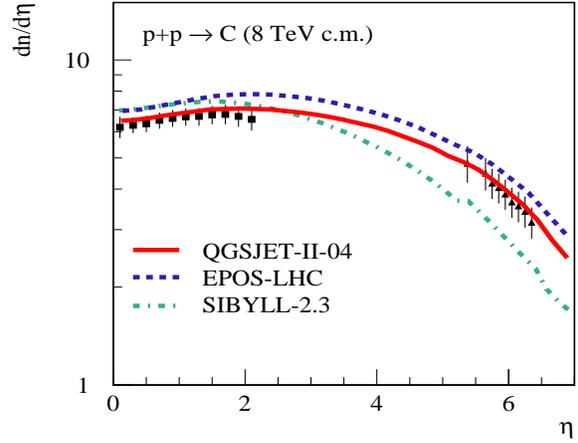}
\caption{$dn^{\rm ch}_{pp}/d\eta$ for $pp$ collisions at $\sqrt{s}=8$ TeV, 
as calculated using the 
 QGSJET-II-04, EPOS-LHC, and SIBYLL-2.3 models 
 (solid, dashed, and dash-dotted lines respectively) for the nondiffractive
 event selection of TOTEM: at least one charged hadron produced
both  at $-6.5<\eta<-5.3$ and at $5.3<\eta<6.5$.   The  CMS and TOTEM data
  are shown  by filled squares and filled triangles respectively.}
\label{fig:cms-tot}       
\end{figure}

The minijet approach of the SIBYLL model is already
disfavored  by
recent combined measurements by the CMS and TOTEM experiments
 of the pseudorapidity $\eta$ density    $dn^{\rm ch}_{pp}/d\eta$ 
 of produced charged hadrons in $pp$ collisions  at $\sqrt{s}=8$ TeV 
  \cite{cms-tot}.  As  one can see in Fig.\ \ref{fig:cms-tot}, %
 $dn^{\rm ch}_{pp}/d\eta$ predicted by
SIBYLL-2.3  steeply falls down at large $\eta$, which
  reflects the quick decrease of the number of constituent partons when 
  parton momentum fraction  increases.
 In contrast,  EPOS-LHC and QGSJET-II-04 predict a much flatter
 $\eta$-dependence for  produced charged hadrons, in a good agreement 
 with the experimental data.

\begin{figure}[t]
\centering
\includegraphics[height=5.9cm,width=0.45\textwidth,clip]{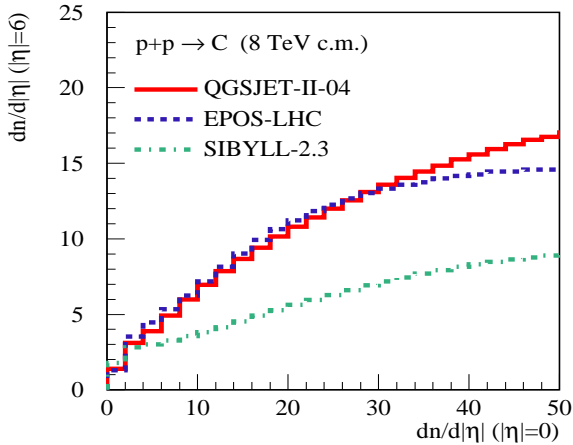}
\caption{Pseudorapidity   density of produced charged hadrons
$dn^{\rm ch}_{pp}/d|\eta|$ at  $|\eta|=6$ ($p_{\rm t}>0$) 
as a function of $dn^{\rm ch}_{pp}/d|\eta|$ at  $|\eta|=0$
 ($p_{\rm t}>0.1$ GeV)
  in $pp$ collisions at $\sqrt{s}=8$ TeV,  as calculated using the 
 QGSJET-II-04, EPOS-LHC, and SIBYLL-2.3 models
 (solid, dashed, and dash-dotted lines respectively).}
\label{fig:cms-tot-corr}       
\end{figure}

However, the crucial discrimination of the minijet approach may be provided
measuring  correlations between the signal strengths in central and 
forward-looking detectors at the LHC.
 For the particular case of the CMS and TOTEM experiments,
  this is illustrated in  Fig.\  \ref{fig:cms-tot-corr}, 
 where we plot for  $\sqrt{s}=8$ TeV 
  $dn^{\rm ch}_{pp}/d|\eta|$  at $|\eta|=6$
  (averaged over the interval $5.5<|\eta|<6.5$) 
  as a function of the central  $\eta$-density ($|\eta|<1$) of
  charged hadrons. Both EPOS-LHC and QGSJET-II-04 predict
a strong correlation of the signal strengths in CMS and TOTEM. The respective
results of the two models practically coincide with each other,
 apart from the tails of the multiplicity distributions.
In contrast, for  SIBYLL~2.3 such a correlation is
twice weaker, being thus a ``smoking gun'' signature for the
desirable discrimination.

Another possible way for the model discrimination is via measurements
 of very forward particle production by the LHCf experiment at the LHC, 
when supplemented by triggering different hadronic activities in   ATLAS,
 as discussed in Ref.\ \cite{ost16}.

\section{RELEVANCE OF PION-AIR INTERACTIONS}
\label{sec:xmax-pions}
The analysis in Section \ref{sec:fock} does not explain the large, up to
40  g/cm$^2$, differences in
  $X_{\max}$ predictions for the other three models 
  which employ essentially the same treatment of 
  constituent parton Fock states.
One might relate these  differences to  present experimental uncertainties 
concerning the rate of the inelastic diffraction in high energy $pp$ collisions.
 Indeed, the diffraction   has a significant influence on the   shape of 
 very forward spectra of secondary hadrons and, through its close
 relation to the inelastic screening effect, on the calculation of the
 inelastic proton-air cross section, which in turn
both make a strong impact on the longitudinal EAS development. 
This has been investigated in Ref.\  \cite{ost14} in the framework of the 
QGSJET-II-04  model.
 The obtained characteristic uncertainty for  $X_{\max}$  amounted to only
 10 g/cm$^2$, while being smaller than 3 g/cm$^2$ for RMS($X_{\max}$).

\begin{figure}[t]
\centering
\includegraphics[height=5.9cm,width=0.45\textwidth,clip]{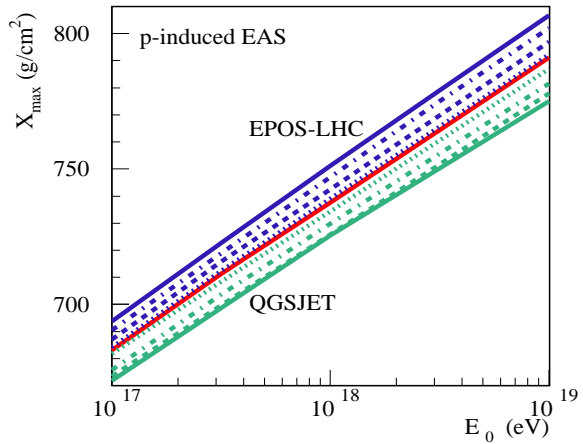}
\caption{Primary energy dependence of  $X_{\max}$
 for $p$-induced vertical EAS, as
calculated using the EPOS-LHC, QGSJET-II-04, and QGSJET models 
(top, middle, and bottom solid lines respectively), 
or applying  mixed model descriptions, 
as explained in the text (dashed, dash-dotted, and dotted lines).}
\label{fig:xmax}       
\end{figure}

To reveal the interaction features which are responsible for the
remaining differences, one can use the ``cocktail'' model approach: 
Using QGSJET-II-04 to describe some selected interactions of hadrons
in air showers or   some particular features of the primary
interaction, while treating the rest with one of the other two models
(see Ref.\  \cite{ost16a} for more details).
As the first step, we apply  QGSJET-II-04 to determine the position of the 
primary particle interaction in the atmosphere and to describe the production
of secondary nucleons in this interaction, which comprises all the effects of 
the inelastic diffraction; all other characteristics of the first
$p$-air  collision and all the subsequent interactions of secondary 
hadrons in the cascade are treated using  EPOS-LHC.
The calculated  $X_{\max}$ values shown by the upper dash-dotted line in 
 Fig.\ \ref{fig:xmax} %
 differ from the original  EPOS-LHC  results by not more than 5 g/cm$^2$,
 which is well within the uncertainty range obtained in Ref.\ \cite{ost14}.
  
  Next, we apply QGSJET-II-04 to describe all the characteristics of the
  primary interaction, while treating the rest of the hadron cascade using 
  EPOS-LHC. The obtained  $X_{\max}$  shown by the upper dashed line in  
   Fig.\ \ref{fig:xmax}    is shifted further towards the  QGSJET-II-04 results
   by up to  5 g/cm$^2$. This additional shift is explained by somewhat harder
   spectra of secondary hadrons in   EPOS-LHC, compared to   QGSJET-II-04. 
 We also repeat the same calculation
 describing secondary hadron interactions in the cascade with the QGSJET model,
 the results being shown by the lower dashed line  in   Fig.\ \ref{fig:xmax}. 
 In this case, the difference with the pure QGSJET-based calculation does not
 exceed 3  g/cm$^2$, which is due to the fact that forward particle spectra
 in proton-air collisions are rather similar in QGSJET and QGSJET-II-04.
 
 There remains a large difference between the two dashed lines in 
  Fig.\ \ref{fig:xmax}, which is entirely due to  the model treatments
 of   high energy pion-air and kaon-air interactions \cite{ost16a}.
The difference between the lower dashed line and the   QGSJET-II-04 results
 is mainly due to the larger pion-air cross section  and     softer
production spectra for secondary mesons in  QGSJET, compared to  QGSJET-II-04. 
The former effect is illustrated by the transition from the lower dashed to
 lower dash-dotted line in  Fig.\ \ref{fig:xmax} while the latter is
 responsible for the difference between the lower dash-dotted and 
 dotted lines in the Figure.

In turn, for EPOS-LHC the remaining difference with the QGSJET-II-04 results
is due to a copious production of baryon-antibaryon pairs 
in pion-air and kaon-air collisions, also due to  harder
(anti-)baryon spectra in  EPOS-LHC \cite{pie08}. This slows down the 
energy dissipation from the hadronic cascade and thus contributes to the
elongation of the  shower profile.
Indeed, if we apply  QGSJET-II-04 to describe
both the primary interaction and the production of nucleons and antinucleons
in all the secondary pion-air and kaon-air collisions,
 while treating the rest with
EPOS-LHC,  the obtained  $X_{\max}$  
shown by the upper dotted line practically
coincides with the  QGSJET-II-04 results.

\section{RELATION TO  MUON PRODUCTION DEPTH}
\label{sec:model-xmu}
As demonstrated in  Section \ref{sec:xmax-pions}, a large part of the model
uncertainty for  the predicted  $X_{\max}$ is related to the
 treatment of pion-air  collisions  at very high energies.
 While there exist no experimental data for such interactions above
  fixed-target energies, one may try to constrain the models 
by studying other EAS characteristics. A particularly promising choice
is the  maximal muon production depth $X^{\mu}_{\max}$ in air showers,
 recently measured by the  Pierre Auger Observatory (PAO)  \cite{pao-xmu}.
 Interestingly,   one observed a strong contradiction between the
respective EPOS-LHC predictions and the experimental data:
The   muon production  maximum was observed
substantially  higher in the atmosphere than predicted by that model 
 for the heaviest possible primary cosmic ray nuclei.

One may generally expect a rather strong sensitivity of the predicted 
$X^{\mu}_{\max}$ to the modeling of pion-air collisions since the EAS
 muon content is formed in a multi-step hadronic cascade in which the
 number of pions and kaons increases in an avalanche way until the probability
 for their decays becomes comparable to the one for interactions. 
 This happens when the   energies for most of pions approach  the  pion
 critical energy, $E_{\rm crit}^{\pi ^{\pm}}\simeq 80$ GeV.
  The position  of the muon production maximum is close to this turning point.

As a consequence,  $X^{\mu}_{\max}$ is very sensitive
to the forward spectral shape of secondary mesons in pion-air collisions:
Producing in each cascade step a meson of a slightly higher energy would mean
that a larger number of cascade branching steps is required for reaching the
critical energy, with the result that the maximum  of the muon production 
profile  will be observed deeper in the atmosphere. On the other hand,
 a smaller pion-air cross section  would increase the pion mean free pass 
 and thus also elongate the muon production profile. Actually, a similar
 effect may be produced by a larger diffractive contribution in pion-air
interactions \cite{pie16}.
In addition, in the particular case of the EPOS model,
its predictions for  $X^{\mu}_{\max}$ may be influenced by
the copious production of  (anti-)baryons in pion-air collisions \cite{pie08}.
Unlike pions and kaons, (anti-)nucleons  participate
in the hadronic cascade without decays, down to the GeV energy range,
producing additional generations of secondary hadrons.
Muons emerging from decays of secondary pions and kaons created in interactions
of  low energy nucleons and antinucleons contribute
 to the elongation of the muon
production profile and give rise to larger values of  $X^{\mu}_{\max}$.

\begin{figure}[t]
\centering
\includegraphics[height=5.9cm,width=0.45\textwidth,clip]{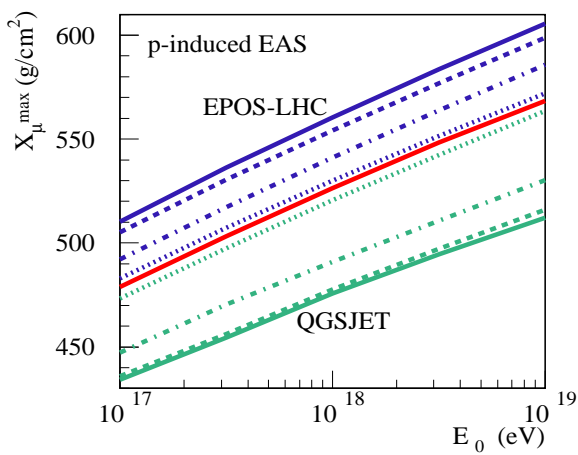}
\caption{Primary energy dependence of  $X^{\mu}_{\max}$
 ($E_{\mu}\geq 1$ GeV) for $p$-induced vertical EAS, as
calculated using the EPOS-LHC, QGSJET-II-04, and QGSJET models 
(top, middle, and bottom solid lines respectively), 
or applying  mixed model descriptions, 
as explained in the text (dashed, dash-dotted, and dotted lines).}
\label{fig:xmumax}       
\end{figure}

Generally, we observe much  larger differences between the model
predictions for  $X^{\mu}_{\max}$, compared
to the case of  $X_{\max}$, as demonstrated in  Fig.\ \ref{fig:xmumax}. %
To reveal the reasons for these differences,
 we  use the same ``cocktail'' model
 approach as in  Section \ref{sec:xmax-pions}. 
 First, we apply QGSJET-II-04 to describe all the characteristics of the
  primary interaction, while treating the rest of the hadronic cascade using 
 either EPOS-LHC or QGSJET, the results shown respectively by the
 upper and lower dashed lines in  Fig.\ \ref{fig:xmumax}. 
  As expected,  the obtained $X^{\mu}_{\max}$ values  deviate only slightly
 from the original model calculations -- as 
  the model predictions for  $X^{\mu}_{\max}$ 
   are dominated by the treatment of secondary
  (mostly pion-air) interactions in the cascade. 
  For example, the smaller $X^{\mu}_{\max}$ values predicted by QGSJET
  are mostly due to somewhat larger inelastic
  pion-air and kaon-air cross sections and softer meson spectra produced
 by that model, compared to QGSJET-II-04. The former effect is illustrated 
  by the difference between the lower dashed and dash-dotted
  lines in  Fig.~\ref{fig:xmumax}, while the latter -- by the difference
  between the lower  dash-dotted and dotted lines in the Figure.

 In turn, a large part of the difference between EPOS-LHC and  QGSJET-II-04
is due to the copious  production of  baryon-antibaryon pairs in 
the former model, as illustrated by the transition from the upper dashed to
the upper dash-dotted line in  Fig.~\ref{fig:xmumax}.  The  remaining
 difference between the two models is mainly due to a larger
  diffractive contribution in pion-air interactions in  EPOS-LHC \cite{pie16}.
Indeed, using  QGSJET-II-04 results both for the primary interaction and for
hadron spectra in   secondary pion-air and kaon-air collisions,
we get the energy-dependence  of  $X^{\mu}_{\max}$, shown by the 
blue dotted line in  Fig.\  \ref{fig:xmumax}, which is very close to 
the pure QGSJET-II-04 calculation.

Thus, we observed that the same features of pion-air interactions, which
had a sizable influence on model predictions for  $X_{\max}$, make a much
stronger impact on the corresponding results for   $X^{\mu}_{\max}$.
This can be used to put strong constraints on the respective model approaches.
In particular, the copious  production of  baryon-antibaryon pairs and the 
large  diffractive contribution in pion-air collisions in  EPOS-LHC are
clearly disfavored by the PAO data.

\section{MODEL PREDICTIONS  FOR RMS($X_{\max}$)}
\label{sec:RMS}
\begin{figure}[t]
\centering
\includegraphics[height=5.9cm,width=0.45\textwidth,clip]{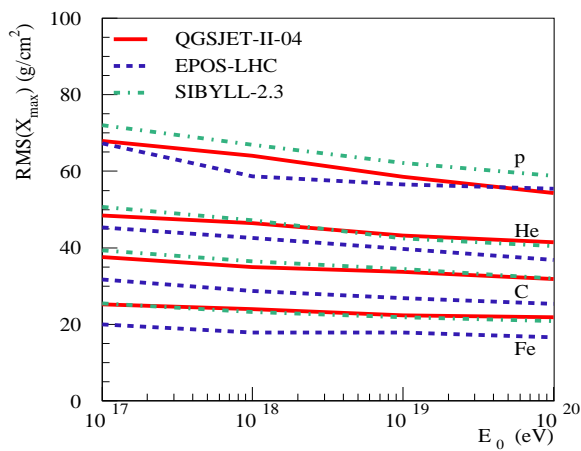}
\caption{Primary energy dependence of  RMS($X_{\max}$)
  for $p$-, He-, C-, and Fe-induced vertical EAS, as
calculated using the QGSJET-II-04, EPOS-LHC, and SIBYLL-2.3 models 
(solid, dashed, and dash-dotted lines respectively).}
\label{fig:rms}       
\end{figure}

As already mentioned above, measurements of the total and elastic proton-proton
cross sections at the LHC strongly constrained model predictions for 
fluctuations of the shower maximum position, RMS($X_{\max}$), 
for proton-induced EAS. The remaining
uncertainties related to the treatment of inelastic diffraction were estimated
to be smaller than 3  g/cm$^2$ \cite{ost14}. Comparing in Fig.\ \ref{fig:rms} 
 the respective results of the interaction models tuned to the LHC data, 
 we observe indeed a very good agreement for the case of the primary proton.
 However,  RMS($X_{\max}$) for nucleus-induced air showers is quite sensitive
 to the treatment of the fragmentation of the nuclear spectator part \cite{qgs93}.
Thus, it is quite remarkable that the corresponding predictions of 
QGSJET-II-04  and SIBYLL-2.3 rather precisely coincide with each other,
despite using different modeling of the  nuclear breakup. In this respect, the
much smaller fluctuations of $X_{\max}$ for nuclear primaries, predicted by
EPOS-LHC, are  surprising.

\begin{figure}[htb]
\centering
\includegraphics[height=5.9cm,width=0.45\textwidth,clip]{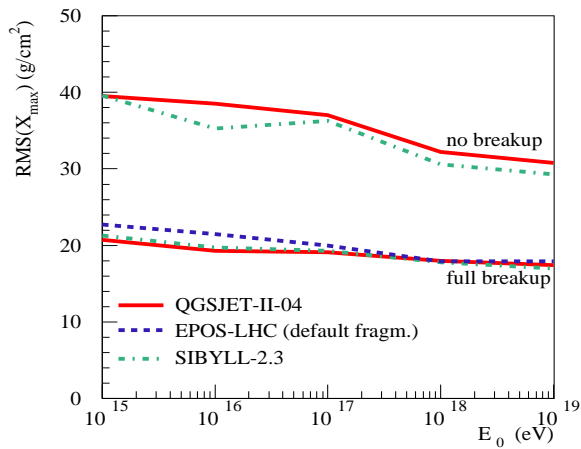}
\caption{Primary energy dependence of  RMS($X_{\max}$) for  Fe-induced
 vertical EAS,  calculated with QGSJET-II-04  and SIBYLL-2.3  
(solid  and dash-dotted lines respectively)
  for two nuclear breakup options, as discussed in the text,
   compared to the default  EPOS-LHC results (dashed line).}
\label{fig:rms-frag}       
\end{figure}

To gain further insight into the issue, we compare in  Fig.\ \ref{fig:rms-frag} 
the results of QGSJET-II-04  and SIBYLL-2.3 for  RMS($X_{\max}$)
of iron-induced EAS, considering two
extreme (and unrealistic) assumptions concerning the nuclear fragmentation: 
Treating the complete spectator part as a single nucleus (no breakup)
or assuming it to disintegrate into separate nucleons (full breakup).
It is easy to see that the latter option gives rise to twice smaller
 $X_{\max}$ fluctuations, compared to the former, as noticed already in
   \cite{qgs93}, and that the predictions of the two models agree rather
    precisely with
   each other, for both scenario. The latter is not surprising since the 
 corresponding results for  RMS($X_{\max}$) are dominated by fluctuations of the
 numbers of ``wounded''  (taking part in the interaction) projectile nucleons,
 which are defined by the geometry of   nuclear collisions in the 
 Glauber-Gribov approach. 
  In contrast, 
 the values of   RMS($X_{\max}$), obtained with EPOS-LHC  
   for the two above-discussed fragmentation options, 
 coincide within 2  g/cm$^2$ with each other and with the default 
 model  results plotted as the dashed line in   Fig.\ \ref{fig:rms-frag}.
 Moreover, the   $X_{\max}$ fluctuations predicted by EPOS-LHC are very close
 to the ones obtained with both  QGSJET-II-04  and SIBYLL-2.3, using the 
full breakup option, which is clearly inconsistent with experimental data
on nuclear fragmentation (see, e.g.,  \cite{fragm} for a review). It is noteworthy
that the above-demonstrated underestimation of  $X_{\max}$ fluctuations 
by EPOS-LHC may bias an analysis of the  cosmic ray mass composition.

\section{FEW COMMENTS ON THE PAO MUON EXCESS AND POTENTIAL SIGNALS OF NEW PHYSICS}
\label{sec:MUON_EXCESS}
Let us finally discuss the outstanding puzzle related to the PAO measurements
of the EAS muon content: For primary energies around  $10^{19}$ eV,
the observed muon density appeared to exceed 
by a large factor ($1.5-2$) the
one predicted by air shower simulations \cite{pao-nmu1,pao-nmu2}.
Potential explanations of such a muon excess are very challenging, keeping
in mind that  the muon content of air showers is formed in a multistep cascade 
process,  which involves hadron-air  interactions over 
a wide energy range. Hence, it is very difficult to create such an excess
at ultra-high energies only, while keeping the simulation results unchanged
at lower energies.

To see that, let us assume that there is no serious problem with the predicted
 EAS muon number $N_{\mu}$ up to primary energies $E_0 \sim 10^{17}$ eV, 
 as indicated by   experimental data
\cite{kascade-nmu1,kascade-nmu2,msu-nmu,ice-nmu}.
Taking into account that most of secondary hadrons are produced with rather
small fractions of the primary particle energy $x_E< 0.1$, there is at most
one cascade step before the secondary particle energies fall below   $10^{17}$ eV,
for a proton-initiated shower of   $10^{19}$ eV. How strong a modification
of the primary interaction is then needed to reproduce the PAO muon excess?
If we assume that some potentially new mechanism allows one to enhance the
multiplicity of the first interaction by as much as a factor of two,
this would result in less than 10\% enhancement of  $N_{\mu}$ at ground level
\cite{ost2006}.

Let us yet further speculate that some new physics emerges around  $10^{19}$ eV,
giving rise to an order of magnitude enhancement of the multiplicity, which 
should just be sufficient to reach an agreement with the PAO results.
For having such a strong effect in average, the new physics should affect
most of the primary collisions, rather than a small fraction of those, i.e.\ to 
emerge  with cross sections comparable to the total inelastic proton-air
cross section at those energies, $\sigma^{\rm inel}_{p-{\rm air}}\sim 0.5$ barn. 
Keeping
in mind that dedicated searches for the Beyond-Standard-Model physics at the
Large Hadron Collider proceed presently at the femtobarn level, this should be
considered as a highly speculative scenario.
It is noteworthy that such an option may be  discriminated
by the Pierre Auger collaboration.
Indeed, based on simple geometry arguments, one can conclude that a 
large contribution to proton-air interactions comes from peripheral
(large impact parameter $b$) collisions, characterized by a small number of
 ``wounded'' target nucleons and relatively
 low parton densities. Such peripheral collisions are thus far from reaching
 extreme conditions for the new physics to emerge. If we then assume that in
 some, say, 10\% of most central (small $b$) interactions a ``hyper-fireball''
 is created, producing an  order of magnitude higher than average multiplicity,
 this would give rise to huge event-by-event fluctuations of the muon density
  at ground, $\delta \rho_{\mu}/\rho_{\mu}>100$\%. Such an order of magnitude
   enhancement of $\delta \rho_{\mu}$ can be easily observed by ground detectors.

\section{OUTLOOK}
\label{sec:Outlook}
Experimental studies of proton-proton collisions at the LHC substantially 
reduced the uncertainties of numerical simulations of CR-induced air showers,
notably, thanks to the precise measurements of the total and elastic $pp$
cross sections by the TOTEM and ATLAS experiments. Nevertheless, there exists
yet a significant spread between the model predictions
for the shower maximum position. The largest between all the models $X_{\max}$
values predicted by SIBYLL-2.3 are due to the very weak energy dependence
of the inelasticity of that model, which is a direct consequence
of   its grounding minijet approach. While that approach is already
disfavored by combined studies of particle production by CMS and TOTEM,
the final discrimination will be provided by measuring correlations between
the signal strengths in central and forward-looking LHC detectors, like
CMS and TOTEM, or ATLAS and LHCf.

Apart from such a discrimination and more precise measurements of the
inelastic diffraction, the LHC potential for improving EAS simulations
is already limited since  the dominant source of uncertainties shifts
 towards the model treatments of very high energy 
 pion-air  interactions. This can be constrained indirectly by studying
 other air shower observables, notably, by measurements
of the   maximal muon production depth. In particular, the PAO results on
  $X^{\mu}_{\max}$   disfavor
   the copious production of   baryon-antibaryon pairs and the large
   diffractive contribution in pion-air collisions,
    predicted by EPOS-LHC. Hence,
    one may expect  that $X_{\max}$  predictions
    of a corrected version    of the model will move closer to the 
     QGSJET-II-04 results.

It is noteworthy that even for QGSJET-II-04 there is a certain tension between
the data of the Pierre Auger experiment on  $X_{\max}$   and  $X^{\mu}_{\max}$
\cite{pao-xmu}:
The latter point towards a heavier CR composition, compared to
the former. In principle, one may try to reach a consistency between the 
two results  by modifying the treatment of pion-air collisions.
 However, as the potential changes would impact  
  $X^{\mu}_{\max}$ much stronger than  $X_{\max}$, this would imply to aim
at higher inelastic cross section and/or softer hadron production spectra
for such interactions, which would push one towards a predominantly light,
presumably proton-dominated CR composition. In turn, this would contradict
other PAO results \cite{pao-rms,pao-corr}.
 Thus, the situation remains confusing and further progress,
both on the experimental and the theoretical sides, is desirable.

\bigskip 
\begin{acknowledgments}
The author  acknowledges the support 
by Deutsche Forschungsgemeinschaft (Project OS 481/1).
\end{acknowledgments}

\bigskip 


\begin{thebibliography}{99}
%
\bibitem{nagano}
M.\ Nagano  and A.\ A.\ Watson, Rev.\ Mod.\ Phys.\ \textbf{72}, 689 (2000).
\bibitem{ung12}
K.-H.\ Kampert and M.\ Unger,
Astropart.\ Phys.\  \textbf{35}, 660 (2012).
\bibitem{wer06}
K.~Werner, F.-M.~Liu, and T.~Pierog,
  Phys.\ Rev.\ C~\textbf{74}, 044902 (2006).
\bibitem{pie15}
T.\ Pierog, Iu.\ Karpenko, J.\ M.\ Katzy, E.\ Yatsenko, and K.\ Werner,
  Phys.\ Rev.\ C~\textbf{92}, 034906 (2015).
 \bibitem{ost06}
 S.~Ostapchenko,   Nucl.~Phys.~Proc.~Suppl.~\textbf{151}, 143 (2006);  
  Phys.~Rev.~D~\textbf{74}, 014026 (2006).
   \bibitem{ost11}
S.\ Ostapchenko,   Phys.\ Rev.\  D \textbf{83}, 014018 (2011);
 EPJ Web Conf.\ \textbf{52},  02001 (2013).
\bibitem{fle94}
R.\ S.~Fletcher, T.\ K.\  Gaisser, P.\ Lipari, and T.\ Stanev,  
  Phys.~Rev.~D~\textbf{50}, 5710 (1994);
E.-J.~Ahn, R.\ Engel, T.~K.\ Gaisser, P.\ Lipari, and T.\ Stanev,
 \textit{ibid.}, \textbf{80}, 094003 (2009).
\bibitem{rie15}
F.\ Riehn, R.\ Engel, A.\ Fedynitch, T.\ K.\ Gaisser, and T.\ Stanev,
Proc.\ Sci., ICRC2015 (2016), 558 [arXiv:1510.00568].
\bibitem{qgs93}
N.~N.~Kalmykov and  S.~S.~Ostapchenko,
Phys.\ Atom.\ Nucl.\ \textbf{56}, 346 (1993).
\bibitem{qgs97}
N.~N.~Kalmykov, S.~S.~Ostapchenko, and A.~I.~Pavlov,
 Nucl.\ Phys.\ Proc.\ Suppl.\ \textbf{52B},  17 (1997).
\bibitem{den11}
D.\ d'Enterria, R.\ Engel, T.\ Pierog, S.\ Ostapchenko, and K.\ Werner,
Astropart.\ Phys.\  \textbf{35}, 98 (2011).
\bibitem{totem13} G.\ Antchev \textit{et al.} (TOTEM Collaboration), 
Europhys.\ Lett.\  \textbf{101}, 21002 (2013);
 \textit{ibid.},   21003 (2013);
 \textit{ibid.},  21004 (2013); 
Phys.\ Rev.\ Lett.\  \textbf{111}, 012001 (2013).
\bibitem{aad14}
G.\ Aad  \textit{et al.} (ATLAS Collaboration),
Nucl.\ Phys.\  B  \textbf{889}, 486 (2014).
\bibitem{pdg}
K.\ Nakamura  \textit{et al.} (Particle Data Group),
 J.\ Phys.\  G \textbf{37},  075021 (2010).
\bibitem{kha15}
V.\ Khachatryan  \textit{et al.} (CMS Collaboration),	
Phys.\  Lett.\  B \textbf{751},  143 (2015).
\bibitem{aab16}
M.\  Aaboud  \textit{et al.} (ATLAS Collaboration),	
 Eur.\ Phys.\ J.\  C \textbf{76},  502 (2016).
\bibitem{ada16}
J.\ Adam  \textit{et al.} (ALICE Collaboration),	
Phys.\  Lett.\  B \textbf{753},  319 (2016).
 \bibitem{cms-tot} S.\ Chatrchyan  \textit{et al.} (CMS and TOTEM
 Collaborations),
 Eur.\ Phys.\ J.\  C \textbf{74}, 3053 (2014).
 \bibitem{ant15} 
G.~Antchev \textit{et al.} (TOTEM Collaboration), 
 Eur.\ Phys.\ J.\  C   \textbf{75}, 126 (2015).
 \bibitem{adr11} 
O.\ Adriani \textit{et al.} (LHCf Collaboration),	
Phys.\  Lett.\  B \textbf{703},  128 (2011);
 \textit{ibid.} \textbf{715},  298 (2012);
 \textit{ibid.} \textbf{750},  360 (2015);
  Phys.~Rev.~D~\textbf{86}, 092001 (2012);
 \textit{ibid.} \textbf{94},   032007 (2016).
 \bibitem{ost16}
 S.~Ostapchenko, M.\ Bleicher,  T.\ Pierog,  and K.~Werner, 
   Phys.\ Rev.\  D \textbf{94},   114026 (2016).
 \bibitem{ulr11}
R.\  Ulrich, R.\ Engel, and M.\ Unger,
 Phys.\ Rev.\  D   \textbf{83}, 054026 (2011).
\bibitem{ost14} S.\ Ostapchenko,   Phys.\ Rev.\  D \textbf{89}, 074009 (2014).
 \bibitem{ost16a}
 S.~Ostapchenko and M.\ Bleicher, 
   Phys.\ Rev.\  D \textbf{93},  051501 (2016).
\bibitem{pie08}
T.\ Pierog and K.\ Werner,
Phys.\ Rev.\ Lett.\  \textbf{101}, 171101 (2008).
\bibitem{pao-xmu}  
A.\ Aab  \textit{et al.} (Pierre Auger Collaboration),
 Phys.\ Rev.\  D \textbf{90}, 012012 (2014);  
 \textit{ ibid.},   \textbf{90},  039904 (2014);  \textit{ibid.},  \textbf{92},  019903
 (2015).
\bibitem{pie16}
T.\ Pierog, 
 EPJ Web Conf.\ \textbf{99}, 09002 (2015);
T.~Pierog, B.\ Guiot, and K.\ Werner,
Proc.\ Sci., ICRC2015 (2016),  337.
\bibitem{fragm}
S.\ Fredriksson, G.\ Eilam, G.\ Berlad, and L.~Bergstr\"om,
  Phys.\ Rep.\ \textbf{144}, 187 (1987).
\bibitem{pao-nmu1}  
A.\ Aab  \textit{et al.} (Pierre Auger Collaboration),
 Phys.\ Rev.\  D \textbf{91}, 032003 (2015);  
 \textit{ ibid.},   \textbf{91},   059901 (2015).
\bibitem{pao-nmu2}  
A.\ Aab  \textit{et al.} (Pierre Auger Collaboration),
 Phys.\ Rev.\  Lett.\ \textbf{117}, 192001 (2016).  
\bibitem{kascade-nmu1}  
W.\ D.\ Apel  \textit{et al.} (KASCADE Collaboration),
Astropart.\ Phys.\  \textbf{24}, 467 (2006).
\bibitem{kascade-nmu2}  
W.\ D.\ Apel  \textit{et al.} (KASCADE-Grande Collaboration),
Astropart.\ Phys.\  \textbf{65}, 55 (2015).
\bibitem{msu-nmu}
  Yu.\ A.\ Fomin \textit{et al.}, arXiv:1609.05764. 
\bibitem{ice-nmu}
J.\ G.\ Gonzalez for the  ICECUBE Collaboration,
J.~Phys.\ Conf.\ Ser.\  \textbf{718}, 052017 (2016).
 \bibitem{ost2006}
 S.~Ostapchenko,  Czech.\ J.\ Phys.\  \textbf{56},  A149 (2006).
\bibitem{pao-rms}  
P.\ Abreu  \textit{et al.} (Pierre Auger Collaboration),
  JCAP \textbf{1302},  026 (2013).  
\bibitem{pao-corr}  
A.\ Aab   \textit{et al.} (Pierre Auger Collaboration),
Phys.\  Lett.\  B \textbf{762},  288 (2016).



\end{thebibliography}
\end{document}